# Antimicrobial Peptide Prediction Using Ensemble Learning Algorithm

Neda Zarayeneh
EECS Department, WSU
Pullman, WA, U.S.
neda.zarayeneh@wsu.edu

Zahra Hanifeloo
EECS Department, ZNU
Strasbourg, France
hanifelo@live.com

*Abstract*—Recently, Antimicrobial peptides (AMPs) have been area of interest in the researches, as the first line of defense against the bacteria. They are raising attention as an efficient way in fighting multi drug resistance. Discovering and identification of AMPs in the wet labs are challenging, expensive, and time consuming. Therefore, using computational methods for AMP predictions have grown attention as they are more efficient approaches. In this paper, we developed a promising ensemble learning algorithm that integrates well-known learning models to predict AMPs. First, we extracted the optimal features from the physicochemical, evolutionary and secondary structure properties of the peptide sequences. Our ensemble algorithm, then trains the data using conventional algorithms. Finally, the proposed ensemble algorithm has improved the performance of the prediction about 10% comparing to the traditional learning algorithms.

## I. Introduction

Bacteria by far are the most diverse, and abundant organisms on Earth. They play an important role in human's life and for decades they have been area of interest in researches [1-3]. Many researches have tried to understand their mechanism by clustering them, find their evolutionary history, or looking at their lateral gene transfer process [3-5]. Most of them have hoped their discoveries might facilitate the perceiving of bacterial antimicrobial-resistant which has become a real threat to global healthcare according to world health organization [6]. Attempts to fighting antimicrobial-resistant has led researchers to a key weapon provided by the nature: Antimicrobial peptides (AMPs).

AMPs, which are suggested to be compelling against microorganisms such as virus, bacteria and fungi, are significant natural immune molecules that establishes a first line of host defense against microorganisms by damaging their cell membrane or their intracellular functions [7].

Developing synthetic anti-microbial drugs can take years, and then antimicrobial resistance always emerges the need for new line of drugs. Because of these obstructions, AMPs have grown attention as an alternative option for conventional approaches [7].

Discovering the AMPs in the wet-labs can be a challenge itself because it is still time consuming. Therefore, with the availability of enough data developing sequence-based computational tools have been found to be an effective way in identifying the peptides with high possibility of being a good AMP candidate [8]. Discovering these types of AMPs prior to the wet-lab experiments increases the probability of designing an AMP in a shorter time [8]. Here we discuss some of the most recent works that have applied the computational biology approaches to predict the AMPs.

In [8], authors have developed a supervised learning algorithm to predict the AMPs. They first extracted physio-chemical and structure-based features, then they trained a Support Vector Machine (SVM) using the input feature.

Their approach increased the accuracy comparing to previous approaches; however, we suggest that accuracy could be increased using ensemble models comparing to a solo SVM.
AMEP [9] is a more recent study that has applied an ensemble learning algorithm to predict the AMPs. Initially, they generated the distribution patterns of amino acids properties as features of the peptides, subsequently they used as input in Random Forest for prediction of AMPs. Their algorithm increased the accuracy comparing to the previous model. However, the precision for their algorithm is not as convincing as the accuracy.

AMAP [10] is another machine learning algorithm developed to predict the antimicrobial activity of the peptides. AMAP has applied multi-label classification to predict several types antimicrobial peptides. They have evaluated their model using cross validation and compared to the state-of-the-art methods, and the result showed improvement in performance.

All the methods mentioned above along with other computational tools listed in [11], have generated useful knowledge for the prediction of AMPs. However, minimizing the number of false positives by improving the algorithm is required. In this study, we made an attempt to develop a computational approach for prediction of antibacterial higher performance. Initially, we generate the features from physiochemical, evolutionary and secondary structure properties of the peptide sequences. Next, we reduce the dimension of the features and finally use them as an input for our ensemble machine learning algorithm. Our approach found to be more accurate than existing approaches. This paper is organized as follows. In section II explain our methodology in detail including data collection, feature extraction, and learning algorithms. Then in section III, we evaluate our approach. Section IV explains the conclusion of our work and future works.

## II. METHODOLOGY

In this section we have explained our methodology in detail. First how we collected the data, then features that we generated, and finally we have discussed the model we built.

### A. Data Collection

We collected positive antibacterial peptides (ABPs) from several publicly available databases. We downloaded in total 5000 positive ABPs available from Data Repository of Antimicrobial Peptides (DRAMP) [12], database Antimicrobial peptides (dbAMP) [13], and Collection of antimicrobial peptides (CAMP) [14]. For the negative dataset, we first computed the average weight of each amino acid in the positive data, and also length distribution of them. Then based on the result we generated 5000 negative peptides with the same weight and length distribution of the positive AMPs.

Figure 1 plots the distribution of the length for positive and negative data, and figure 2 represents the distribution of the positive and negative data in terms of grand average of hydropathicity (gravy), and molecular weight of the sequences. The plots show how close the generated negative datasets are to the positive peptides. Using such a stringent dataset will affirm the result of model.

### B. Feature Extraction

We extracted different features for the peptide sequences. We searched through recent researches to find the optimal features. A number of researches [15-17] have suggested using physicochemical, evolutionary and secondary structure properties as optimal features for the peptides.

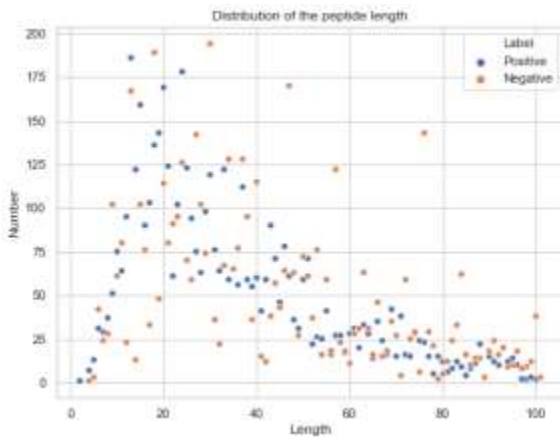

Figure 1- The distribution of the positive and negative AMPs in terms of the lengths and number in the dataset.

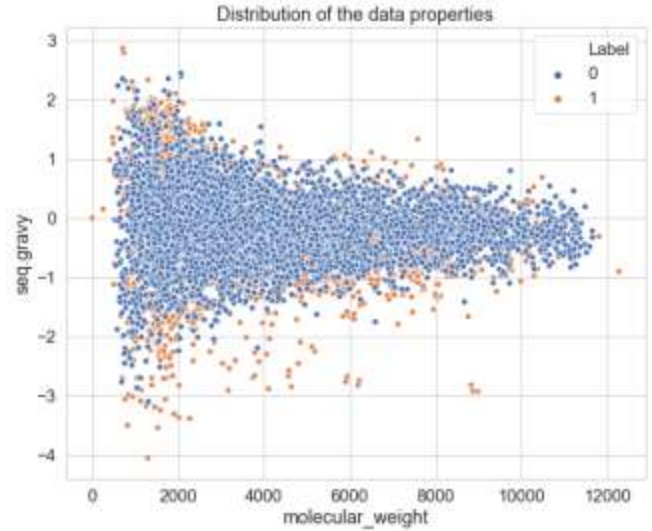

Figure 2- The distribution of the positive and negative AMPs in terms of sequence grand average of hydropathicity (gravy), and the molecular weight of the sequence

Table 1 lists the features that have been generated. Amino acid decomposition for each sequence is a fraction of the amino acids to the peptide length. The composition, transition, and distribution (CTD) model examines the physicochemical properties of the amino acids such as normalized van der Waals volume, hydrophobicity, polarity, polarizability, and secondary structure. There are 591 feature per sequence for these three feature sets. iFeature [17] is a python-based tool that has implemented the code for most of the protein sequence features. We used the classes developed by iFeature and also [15]. In order to mitigate the number of features, we first computed the Pearson's correlation coefficient (1) between the features.

$$Pearson(A, B) = \frac{E((A - \mu_A)(B - \mu_B))}{\sigma_A \sigma_B} \quad (1)$$

Where $E$ is the expectation, and $\mu_A$ and $\mu_B$ are the mean values, and $\sigma_A$ and $\sigma_B$ are the standard deviations of A and B, respectively. The result of correlation is a number between [-1, +1]. The farther from zero indicates the higher correlation between A, and B. Here, we kept the features with |correlation|<0.90. This way we reduced the number of features from 591 to 49.

Table 1- Features for the peptides

| Feature | Dimension |
|---|---|
| amino acid composition | 20 |
| composition, transition, and distribution (CTD) model | 168 |
| Predicted secondary structure | 3 |
| position-specific scoring matrix (PSSM) | 400 |

## C. Learning Algorithm

We trained our model using three well-known machine learning algorithms: Support Vector Machine (SVM) [18], Random Forest (RF) [19], Gradient Boost Model (GBM) [20]. Then we developed an ensemble [15] algorithm that utilizes the learning by combining the three algorithms.

### 1) SVM

Support Vector Machine (SVM) is a non-probabilistic, linear, binary classifier that can be used for both regression and classifying data by learning a hyperplane which divides the classes of the data. SVM basically learns an (n – 1)- dimension hyper plan for an n-dimensional space into two classes. SVM can be also used for classifying a non-linear dataset by projecting the dataset into a higher dimension in which it is linearly separable. It has low performances when the data is noisy.

### 2) Randome Forest

Random forest [19] is a well-known ensemble algorithm that works by combining a large number of decision trees. The RF algorithm operates by voting. It simply benefits from the wisdom of the crowd. Every individual tree in the random forest predicts a class for the datapoint and the class with the highest number of votes turns into the final prediction. Training a large number of the uncorrelated decision trees is the key that RF works well. Uncorrelated trees lead to a higher accurate prediction, and also the trees protect each other from their individual errors. For building a random forest model, the features and in result the trees generated based on those features are required to have low correlation.

### 3) Gradiant Boost Model

Gradient boosting is another ensemble learning algorithm that predictors are not independent, and they work sequentially. The gradient boosting algorithm (GBM) is basically a technique for both regression and classification problems. It generates a prediction model in the form of an ensemble of weak prediction models, typically decision trees. It builds the model in a stage-wise fashion like other boosting methods do, and it generalizes them by allowing optimization of an arbitrary differentiable loss function.

### 4) Ensemble Method

We generated an Ensemble learning algorithm using RF, GBM and SVM. As we can see in figure 3, first the base classifiers (RF, GBM, and SVM), take the training dataset as input, then they provide a decision individually. We have mapped the categorical labels "positive" and "negative" to 1, and 0 respectively. Let's the output of their decision be $O_{RF}$, $O_{GBM}$, and $O_{SVM}$, then the final decision is calculated as follows (2).

$$f = \frac{O_{RF} + O_{GBM} + O_{SVM}}{3} \quad (2)$$

$$\text{if} \begin{cases} f == 1 & \rightarrow \text{Strong Positive} \\ f >= 0.66 & \rightarrow \text{Positive} \\ f <= 0.33 & \rightarrow \text{Negative} \\ f == 0 & \rightarrow \text{Strong Negative} \end{cases} \quad (3)$$

The final decision is made based on the result of the $f$. We are able to make suggestion about the probability of being positive or negative based on the result of $f$. However, the final decision for classifying into two classes is if $f>0.5$ the prediction is positive otherwise it is negative.

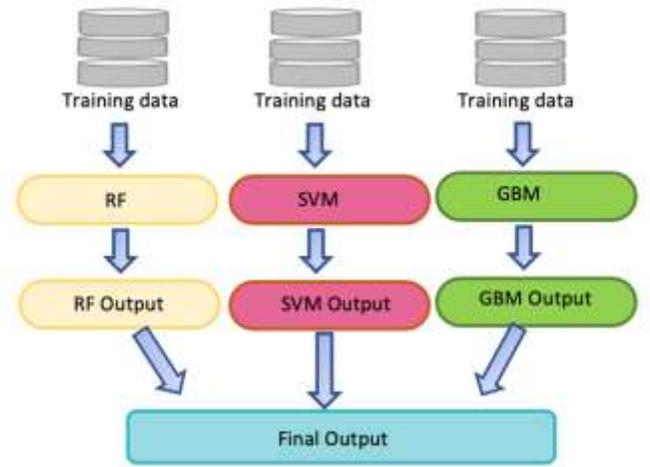

Figure 3- Ensemble method created by RF, GBM, and SVM

## III. RESULTS

For evaluating our model, we used four different evaluation metrics [21]: Accuracy (4), F1 Score (5), Recall (6), and ROC (7). First we define True Positives (TP), True Negatives (TN), False Positives (FP) and False Negatives (FN).
TPs are the peptides correctly predicted as antibacterial peptides. TNs are peptides that are correctly predicted as not antibacterial peptides. FPs occur when a not antibacterial peptide is predicted as antibacterial. FNs happens when the predicted value indicates the peptide is not antibacterial, while the actual value is antibacterial peptide. The evaluation metrics are defined based on these parameters.

$$Accuracy = \frac{TP + TN}{TP + TN + FP + FN} \quad (4)$$

$$Recall = \frac{TP}{TP + FN} \quad (5)$$

$$F1\ Score = \frac{2TP}{2TP + FP + FN} \quad (6)$$

$$FPR = \frac{FP}{FP + TN} \quad (7)$$

$$TPR = \frac{TP}{FN + TP} \quad (8)$$

The Receive Operating Characteristic (ROC) curve is the created by plotting TPR against FPR. It shows the ability of the model to classify a binary dataset.

We hold out 25 percent of the data as test, and trained the model using the 75% of the data. Table 2 compares the performance result for the ensemble method and three individual models.

**Table 2- Performance Evaluation**

| Method | Accuracy | F1 Score | Recall |
|---|---|---|---|
| SVM | 0.75 | 0.73 | 0.69 |
| GBM | 0.63 | 0.61 | 0.58 |
| RF | 0.76 | 0.76 | 0.74 |
| Ensemble | 0.87 | 0.86 | 0.86 |

The table shows that generally there is almost 10 percent improvement in prediction using the ensemble method. The higher F1 score means, the ensemble has been able to improve the precision of the model.

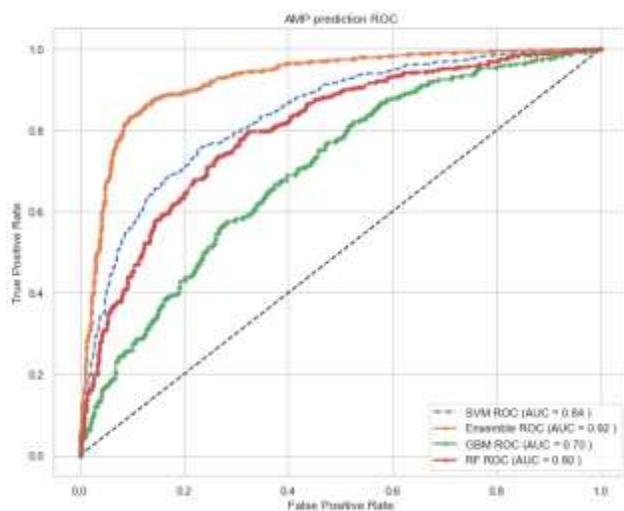

Figure 4 – The ROC curve for the proposed ensemble method and for three other individual learning algorithms

Figure 4 plots the ROC curve for the ensemble algorithm and three individual algorithms. The figure shows that all the models are better than random selecting the peptides. SVM works better than other two models. The Ensemble model benefits from combining the three models, and the higher area under curve (AUC) shows the improvement.

## IV. CONCLUSION

Recently, predicting antimicrobial peptides has grown attention. In this work we developed a learning algorithm to predict the antibacterial peptides. The contribution of our work comes from combining well-know algorithms to generate a more powerful learning algorithm. We trained and tested our results using a highly stringent data, and the result shows almost 10% performance improvement. For the future work, we will design an ensemble model for predicting all types of antimicrobial peptides. Also, we will try to design a meta classifier to improve our model even more.